\documentclass[aps,prl,numerical,superscriptaddress,showpacs,floatfix,reprint]{revtex4-1}

\usepackage{graphicx,color}
\usepackage{amsmath,amssymb,amsfonts}
\usepackage[colorlinks=true,linkcolor=blue,plainpages=false]{hyperref}

\def \be {\begin{equation}}
\def \ee {\end{equation}}
\def \ee  {\end{equation}}
\def \bea {\begin{eqnarray}}
\def \eea {\end{eqnarray}}

\def \l   {\lambda}

\def \roots{\sqrt{s_{_{NN}}}}

\newcommand{\pT} {\ensuremath{p_{\mathrm{T}}}}

\def \lt {\mbox{$<$}}

\begin{document}

\title{Azimuthal harmonics in small and large collision systems at RHIC top energies}

\affiliation{Abilene Christian University, Abilene, Texas   79699}
\affiliation{AGH University of Science and Technology, FPACS, Cracow 30-059, Poland}
\affiliation{Alikhanov Institute for Theoretical and Experimental Physics, Moscow 117218, Russia}
\affiliation{Argonne National Laboratory, Argonne, Illinois 60439}
\affiliation{Brookhaven National Laboratory, Upton, New York 11973}
\affiliation{University of California, Berkeley, California 94720}
\affiliation{University of California, Davis, California 95616}
\affiliation{University of California, Los Angeles, California 90095}
\affiliation{University of California, Riverside, California 92521}
\affiliation{Central China Normal University, Wuhan, Hubei 430079 }
\affiliation{University of Illinois at Chicago, Chicago, Illinois 60607}
\affiliation{Creighton University, Omaha, Nebraska 68178}
\affiliation{Czech Technical University in Prague, FNSPE, Prague 115 19, Czech Republic}
\affiliation{Technische Universit\"at Darmstadt, Darmstadt 64289, Germany}
\affiliation{E\"otv\"os Lor\'and University, Budapest, Hungary H-1117}
\affiliation{Frankfurt Institute for Advanced Studies FIAS, Frankfurt 60438, Germany}
\affiliation{Fudan University, Shanghai, 200433 }
\affiliation{University of Heidelberg, Heidelberg 69120, Germany }
\affiliation{University of Houston, Houston, Texas 77204}
\affiliation{Indiana University, Bloomington, Indiana 47408}
\affiliation{Institute of Modern Physics, Chinese Academy of Sciences, Lanzhou, Gansu 730000 }
\affiliation{Institute of Physics, Bhubaneswar 751005, India}
\affiliation{University of Jammu, Jammu 180001, India}
\affiliation{Joint Institute for Nuclear Research, Dubna 141 980, Russia}
\affiliation{Kent State University, Kent, Ohio 44242}
\affiliation{University of Kentucky, Lexington, Kentucky 40506-0055}
\affiliation{Lawrence Berkeley National Laboratory, Berkeley, California 94720}
\affiliation{Lehigh University, Bethlehem, Pennsylvania 18015}
\affiliation{Max-Planck-Institut f\"ur Physik, Munich 80805, Germany}
\affiliation{Michigan State University, East Lansing, Michigan 48824}
\affiliation{National Research Nuclear University MEPhI, Moscow 115409, Russia}
\affiliation{National Institute of Science Education and Research, HBNI, Jatni 752050, India}
\affiliation{National Cheng Kung University, Tainan 70101 }
\affiliation{Nuclear Physics Institute AS CR, Prague 250 68, Czech Republic}
\affiliation{Ohio State University, Columbus, Ohio 43210}
\affiliation{Institute of Nuclear Physics PAN, Cracow 31-342, Poland}
\affiliation{Panjab University, Chandigarh 160014, India}
\affiliation{Pennsylvania State University, University Park, Pennsylvania 16802}
\affiliation{Institute of High Energy Physics, Protvino 142281, Russia}
\affiliation{Purdue University, West Lafayette, Indiana 47907}
\affiliation{Pusan National University, Pusan 46241, Korea}
\affiliation{Rice University, Houston, Texas 77251}
\affiliation{Rutgers University, Piscataway, New Jersey 08854}
\affiliation{Universidade de S\~ao Paulo, S\~ao Paulo, Brazil 05314-970}
\affiliation{University of Science and Technology of China, Hefei, Anhui 230026}
\affiliation{Shandong University, Qingdao, Shandong 266237}
\affiliation{Shanghai Institute of Applied Physics, Chinese Academy of Sciences, Shanghai 201800}
\affiliation{Southern Connecticut State University, New Haven, Connecticut 06515}
\affiliation{State University of New York, Stony Brook, New York 11794}
\affiliation{Temple University, Philadelphia, Pennsylvania 19122}
\affiliation{Texas A\&M University, College Station, Texas 77843}
\affiliation{University of Texas, Austin, Texas 78712}
\affiliation{Tsinghua University, Beijing 100084}
\affiliation{University of Tsukuba, Tsukuba, Ibaraki 305-8571, Japan}
\affiliation{United States Naval Academy, Annapolis, Maryland 21402}
\affiliation{Valparaiso University, Valparaiso, Indiana 46383}
\affiliation{Variable Energy Cyclotron Centre, Kolkata 700064, India}
\affiliation{Warsaw University of Technology, Warsaw 00-661, Poland}
\affiliation{Wayne State University, Detroit, Michigan 48201}
\affiliation{Yale University, New Haven, Connecticut 06520}

\author{J.~Adam}\affiliation{Creighton University, Omaha, Nebraska 68178}
\author{L.~Adamczyk}\affiliation{AGH University of Science and Technology, FPACS, Cracow 30-059, Poland}
\author{J.~R.~Adams}\affiliation{Ohio State University, Columbus, Ohio 43210}
\author{J.~K.~Adkins}\affiliation{University of Kentucky, Lexington, Kentucky 40506-0055}
\author{G.~Agakishiev}\affiliation{Joint Institute for Nuclear Research, Dubna 141 980, Russia}
\author{M.~M.~Aggarwal}\affiliation{Panjab University, Chandigarh 160014, India}
\author{Z.~Ahammed}\affiliation{Variable Energy Cyclotron Centre, Kolkata 700064, India}
\author{I.~Alekseev}\affiliation{Alikhanov Institute for Theoretical and Experimental Physics, Moscow 117218, Russia}\affiliation{National Research Nuclear University MEPhI, Moscow 115409, Russia}
\author{D.~M.~Anderson}\affiliation{Texas A\&M University, College Station, Texas 77843}
\author{R.~Aoyama}\affiliation{University of Tsukuba, Tsukuba, Ibaraki 305-8571, Japan}
% ...

\collaboration{STAR Collaboration}\noaffiliation

\date{\today}

\begin{abstract}
The first ($v_1^{\text{even}}$), second ($v_2$) and third ($v_3$) harmonic coefficients of the azimuthal particle distribution at mid-rapidity, 
are extracted for charged hadrons and studied as a function of transverse momentum ($p_T$) and mean charged particle multiplicity density 
$\langle \mathrm{N_{ch}} \rangle$  in U+U ($\roots =193$~GeV), Au+Au, Cu+Au, Cu+Cu, $d$+Au and $p$+Au collisions 
at $\roots = 200$~GeV with the STAR Detector.    
For the same $\langle \mathrm{N_{ch}} \rangle$, the $v_1^{\text{even}}$ and $v_3$ coefficients are observed to be 
independent of collision system, while $v_2$ exhibits such a scaling only when  normalized by the initial-state 
eccentricity ($\varepsilon_2$). The data also show that $\ln(v_2/\varepsilon_2)$ scales linearly 
with $\langle \mathrm{N_{ch}} \rangle^{-1/3}$. These measurements provide insight into initial-geometry 
fluctuations and the role of viscous hydrodynamic attenuation on $v_n$ from small to large collision systems.                                                                                                   

\end{abstract}
\maketitle


An important goal of the experimental program at the Relativistic Heavy Ion Collider (RHIC) 
at Brookhaven National Laboratory (BNL), is to provide quantitative experimental data which can
(i) give insight on the dynamical evolution of the quark-gluon plasma (QGP) created in heavy ion collisions, 
and (ii) serve as important constraints for the extraction of the associated transport coefficients.” 
 The azimuthal anisotropy of particle emission in the transverse plane, known as anisotropic flow, is a key observable 
because it reflects the viscous hydrodynamic response to the initial spatial distribution 
in energy density (both from intrinsic geometry and fluctuations), produced in 
the early stages of the collision 
\cite{Danielewicz:1998vz,Ackermann:2000tr,Adcox:2002ms,Heinz:2001xi,Hirano:2005xf,Huovinen:2001cy,Hirano:2002ds,Romatschke:2007mq,Luzum:2011mm,
Song:2010mg,Qian:2016fpi,Schenke:2011tv,Teaney:2012ke,Gardim:2012yp,Lacey:2013eia}.

Experimentally, anisotropic flow manifests as an azimuthal asymmetry of the measured single-particle 
distribution, quantified by the complex flow coefficients \cite{Bilandzic:2010jr,Luzum:2011mm,Teaney:2012ke}: 
%
\begin{equation}
 V_n  \equiv v_ne^{in\Psi_n} = \lbrace e^{in\phi}\rbrace_{j},
\label{Vndef}
\end{equation}
where $\phi$ denotes the azimuthal angle around the beam direction, of a particle 
emitted in the collision and $\lbrace \rbrace_{j}$ denotes the average over all particles emitted 
in the event. The weighted average ${{\left< \left| V_n \right|^2 \right>}^{1/2} = v_n }$ (which accounts for 
multiplicity variations) and ${\Psi_n}$ denote the magnitude and azimuthal direction of 
the $n^{\text{th}}$-order harmonic flow vector that fluctuates from event to event. 
The first three coefficients, ${v_1}$, ${v_2}$, and ${v_3}$,  
are termed directed, elliptic, and triangular flow, respectively. The fluctuations-driven 
component of ${v_1}$, termed ${v^{\text{even}}_{1}}$, is proportional to the   
dipole asymmetry of the collision system \cite{Teaney:2010vd,Luzum:2010fb}.

The ${v_n}$ coefficients are also related to the Fourier coefficients ${v_{nn}}$ 
which characterize the amplitude of  the two-particle correlations in relative azimuthal angle 
$\Delta\phi=\phi_{\mathrm{a}}-\phi_{\mathrm{b}}$~\cite{Poskanzer:1998yz,Lacey:2005qq} for the particles $\mathrm{a}$ 
and $\mathrm{b}$, which comprise the pairs:
\begin{eqnarray}
\label{eq:2}
{\frac{dN^{pairs}}{d\Delta\phi}\propto1+2\sum_{n=1}^{\infty}v_{nn}\cos(n\Delta\phi)}, \nonumber \\
{v_{nn}(\pT^{a},\pT^{b})  = {v_n(\pT^{a})v_n(\pT^{b})+ \delta_{NF}}},
\end{eqnarray}
where $\delta_{\text{NF}}$ signify the contributions of  short-range non-flow correlations due to 
resonance decays, Bose-Einstein correlations and jet-like decays, as well as long-range contributions, 
which result from global momentum conservation \cite{Lacey:2005qq,Luzum:2010fb,Retinskaya:2012ky,ATLAS:2012at}.

The initial anisotropic density profile $\rho_e(r,\varphi)$ in the transverse ($\perp$) plane, 
which drives anisotropic flow, can be similarly characterized by complex eccentricity 
coefficients \cite{Alver:2010dn,Petersen:2010cw,Lacey:2010hw,Teaney:2010vd,Qiu:2011iv}:
\begin{equation}
{\mathcal{E}_n  \equiv \varepsilon_ne^{in\Phi_n}  =
  - \frac{\int d^2r_\perp\, r^m\,e^{in\varphi}\, \rho_e(r,\varphi)}
           {\int d^2r_\perp\, r^m\,\rho_e(r,\varphi)}},                                                      
\label{epsdef1}
\end{equation}
where ${\Phi_n}$ is the angle of the so-called ${n^{th}}$-order participant plane;
${m={n}}$ for ${n{\geq\,}2}$ and ${m= 3}$ for ${n= 1}$ \cite{Teaney:2010vd}. 
Theoretical investigations show that ${v_n \propto \varepsilon_n}$ 
for  elliptic and triangular flow (${n=2,3}$) 
\cite{Qiu:2011iv,Fu:2015wba,Niemi:2015qia,Noronha-Hostler:2015dbi}, 
and the temperature ($T$) dependent specific shear viscosity $\frac{\eta}{s}(T)$, of the created medium, reduces the ratio 
${v_n/\varepsilon_n}$. Thus, the comparison of viscous hydrodynamical model calculations 
to this ratio is commonly employed to estimate $\frac{\eta}{s}(T)$ and its average $\left<\frac{\eta}{s}(T)\right>$,
over the system's evolution    
\cite{Hirano:2005xf,Romatschke:2007mq,Song:2010mg,Schenke:2010rr,Bozek:2010wt,
Qiu:2011iv,Schenke:2011tv,Niemi:2012ry,Gardim:2012yp,McDonald:2016vlt,Bernhard:2016tnd}.
The viscous attenuation of ${v_n/\varepsilon_n}$ can also be understood within an acoustic model 
framework, akin to that for viscous relativistic hydrodynamics \cite{Liu:2018hjh,Liu:2018xae,Lacey:2013is,Magdy:2018ufy,Magdy:2017kji,Shuryak:2013ke,Lacey:2011ug}: 
\begin{eqnarray}
{\ln({v_{n}}/{\varepsilon_{n}}) \propto - n^{2} \left< \frac{\eta}{s}(T)\right> \langle N_{ch} \rangle ^{-1/3}},
\label{eq:4}
\end{eqnarray}
%
where $\mathrm{\langle N_{ch} \rangle}$ is the charged particle multiplicity density and $\mathrm{\langle N_{ch} \rangle ^{-1/3}}$ is a proxy for the 
dimensionless size of the system \cite{Liu:2018hjh,Liu:2018xae,Lacey:2016hqy}.

 Recent measurements at both RHIC and the Large Hadron Collider (LHC), have indicated  sizable 
${v_2}$ and ${v_3}$ values in high multiplicity $p+p$~\cite{Khachatryan:2015lva,Aad:2015gqa},  
$d$+Au~\cite{Adare:2014keg,Adamczyk:2014fcx} and $p$+Pb collisions \cite{CMS:2012qk,Abelev:2012ola,Aad:2012gla}, 
reminiscent of those observed in peripheral A+A collisions. These measurements have generated  
considerable debate on whether the final-state collective effects, which dominate the mechanism 
for anisotropic flow in A+A collisions, also drive the anisotropy measured in 
high-multiplicity $p+p$ and $p$+A ($d$+A) collisions \cite{Dusling:2013qoz,Bozek:2013uha,Dusling:2015gta}. 
The related question of whether the  properties of the medium produced in the 
small $p+p$, $p$+A and $d$+A systems are similar to those produced in the 
larger A+A systems is also not fully settled.

In this letter we present and compare a comprehensive set of ${v^{\text{even}}_{1}}$, ${v_{2}}$ and ${v_{3}}$ 
measurements for U+U ($\roots = 193$~GeV), Au+Au, Cu+Cu, Cu+Au, $d$+Au, and  $p$+Au collisions at $\roots = 200$~GeV, that should 
prove invaluable for the interpretation of collectivity in small systems, and in ongoing efforts to constrain theoretical models and 
obtain a robust extraction of $\frac{\eta}{s}(T)$. 

%
%
 \begin{figure*}[t]
 \centering{
\includegraphics[width=0.65\linewidth,angle=0.0]{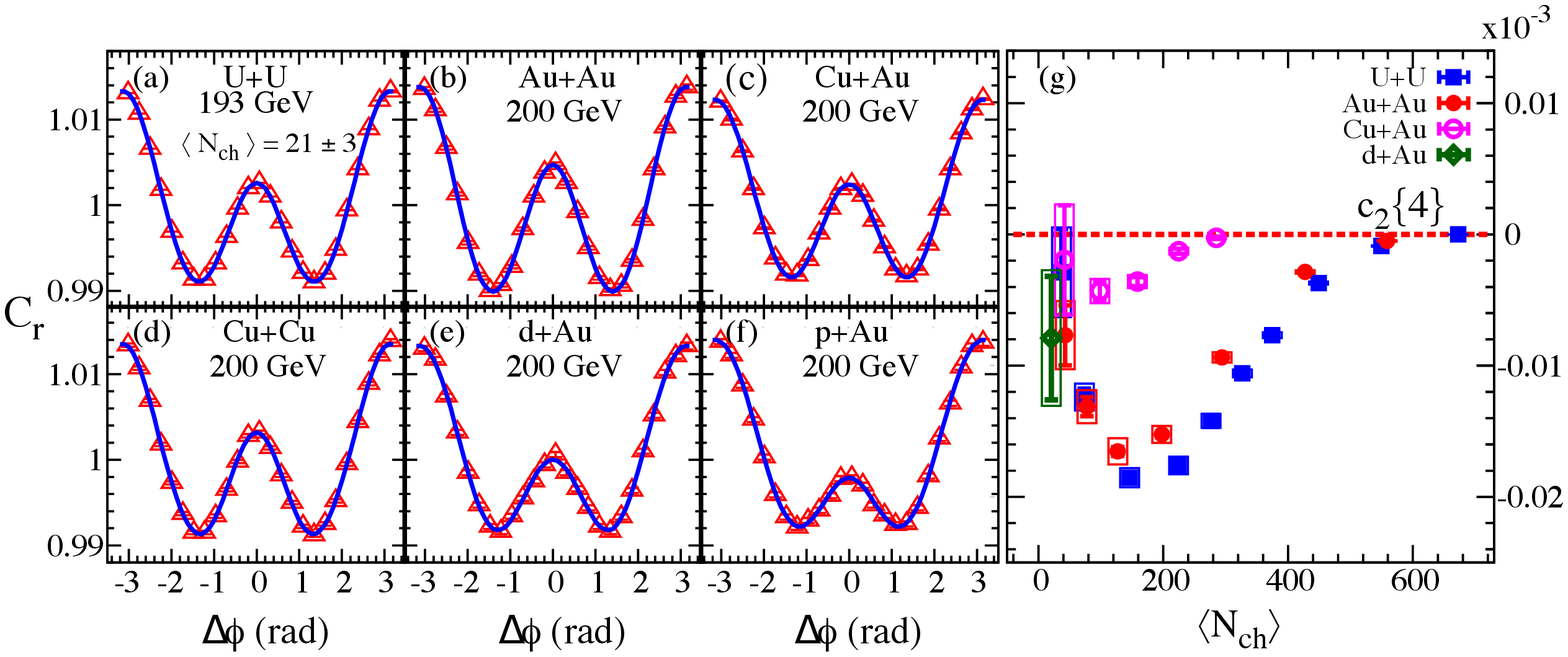}
\vskip -0.35cm
\caption{ Two-particle azimuthal correlation functions (a-f) and four-particle cumulants (g) for $\pT$-integrated 
track pairs ($-1\alt \eta \alt 1$). Results are shown for U+U (a) collisions ($\roots = 193$~GeV) 
and  Au+Au (b), Cu+Au (c), Cu+Cu (d), $d$+Au (e) and $p$+Au (f) collisions ($\roots = 200$~GeV) 
for $\langle \mathrm{N_{ch}} \rangle = 21 \pm 3$. The solid curves show the result of a Fourier fit 
to the data. Panel (g) shows the second-order cumulant $c_{2}\lbrace 4 \rbrace$ vs. $\langle \mathrm{N_{ch}} \rangle$, 
obtained from the same data sets for the systems indicated. \label{corr-func}
 }
}
\vskip -0.2cm
\end{figure*}
The data for the six colliding systems presented in this work, were  collected with the STAR detector at RHIC using a minimum-bias trigger. Charged-particle tracks, measured in the full azimuth and pseudorapidity range ($|\eta|<1.0$) of the Time Projection Chamber (TPC)~\cite{Anderson:2003ur}, were used to reconstruct the collision vertices. Events were selected with vertex positions $\pm 30$~cm from the nominal center of the TPC (in the beam direction).  

Collision centrality and the associated $\langle \mathrm{N_{ch}} \rangle$ was determined from the measured event-by-event multiplicity with the aid of a tuned Monte Carlo Glauber calculation~\cite{Adamczyk:2012ku}. Analyzed tracks were required to have a distance of closest approach to the primary vertex of less than 3~cm, and have at least 15 TPC space points used in their reconstruction. To remove split tracks, the ratio of the number of fit points to a maximum possible 
number of TPC space points was required to be larger than~0.52. Analyzed tracks were restricted to $0.2<\pT<4$~GeV/$c$.

Two-particle $\Delta\phi$ correlation functions ($C_{\text{r}}$) were generated to extract the flow coefficients:
\begin{eqnarray}\label{corr_func}
 C_{\text{r}}(\Delta\phi, \Delta\eta) = \frac{(dN/d\Delta\phi)_{\text{same}}}{(dN/d\Delta\phi)_{\text{mixed}}},
\end{eqnarray} 
where  $(dN/d\Delta\phi)_{\text{same}}$ represents the distribution of  track pairs, 
in relative azimuthal angle $\Delta\phi$, taken from the same event. $(dN/d\Delta\phi)_{\text{mixed}}$ 
represents the $\Delta\phi$ distribution for track pairs in which each member  
is selected from different events in the same $\langle \mathrm{N_{ch}} \rangle$ 
and vertex position classes. The pseudorapidity requirement  $|\Delta\eta| > 0.7$ was imposed 
for all track pairs to suppress short-range non-flow contributions \cite{Star:2018zpt}. A further check for 
the dominance of flow correlations was obtained by measuring the 
second-order four-particle cumulant $c_{2}\lbrace 4 \rbrace$:
\begin{eqnarray}
c_{2}\lbrace 4 \rbrace = \langle\langle 4 \rangle\rangle - 2 \langle\langle 2 \rangle\rangle^{2},
\end{eqnarray}
where $\langle\langle  \rangle\rangle$ represents the averaging first over particles in an event and 
then over all events within a given event class. The three sub-events method \cite{Jia:2017hbm} 
was used for these evaluations with sub-events for $\eta_{1} < -0.35$, $|\eta_{2}| < 0.35$ and $\eta_{3} > 0.35$.

 \begin{figure*}[t]
\centering{
\includegraphics[width=0.60\linewidth,angle=0.0]{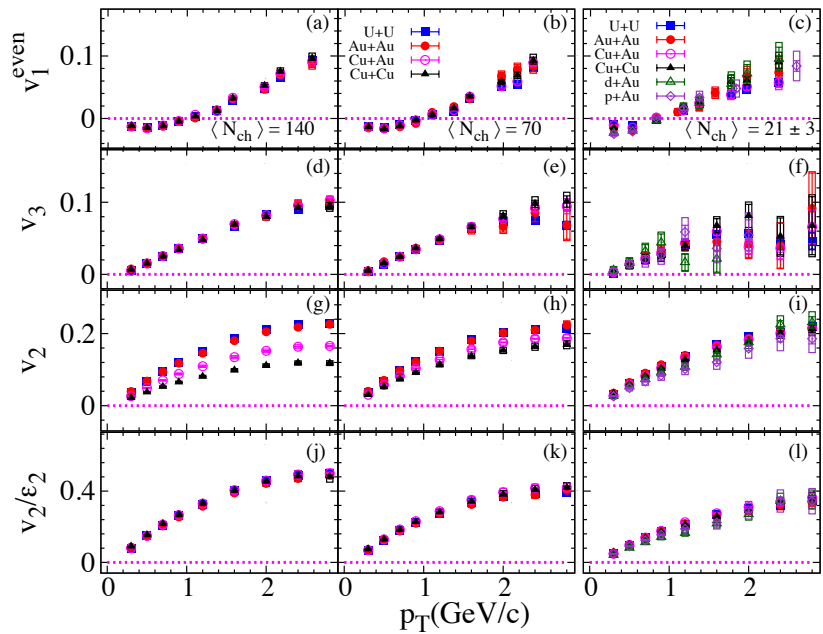}
\vskip -0.35cm
\caption{${v^{\text{even}}_{1}}$, ${v_{2}}$, ${v_{3}}$ and 
${v_{2}/\varepsilon_{2}}$  vs. ${\pT}$ for several $\langle {N_{ch}}\rangle$ 
selections. Results are compared for U+U, Au+Au, Cu+Au and Cu+Cu  for 
$\langle \mathrm{N_{ch}} \rangle = 140$, and $\langle \mathrm{N_{ch}} \rangle = 70$ 
and for U+U, Au+Au, Cu+Au, Cu+Cu, $d$+Au and $p$+Au  for $\langle \mathrm{N_{ch}} \rangle = 21\pm 3$.
} \label{vn:pt}
}
\vskip -0.4cm
\end{figure*}

Figures~\ref{corr-func}(a-e) show the correlation functions obtained for U+U, Au+Au, Cu+Au, Cu+Cu, $d$+Au and $p$+Au 
collisions for $\langle \mathrm{N_{ch}} \rangle = 21 \pm 3$. They indicate patently similar correlation patterns 
with  a visible enhancement of near-side ($\Delta\phi \sim 0$) pairs, reminiscent of the so-called ``ridge" 
observed in high multiplicity $p+p$~\cite{Khachatryan:2015lva,Aad:2015gqa}, $d$+Au~\cite{Adare:2014keg,Adamczyk:2014fcx} 
and $p$+Pb collisions \cite{Aad:2012gla,CMS:2012qk}. The corresponding values for 
$c_{2}\lbrace 4 \rbrace$ vs. $\langle \mathrm{N_{ch}} \rangle$, shown in Fig.~\ref{corr-func} 
(g), indicate negative values which suggests the absence of significant short-range 
non-flow contributions, and the dominance of flow correlations to $C_{\text{r}}$ \cite{Borghini:2001vi,Zhao:2017rgg}.
Note that the paucity of central $p$+Au events precluded the extraction of $c_{2}\lbrace 4 \rbrace$ from 
these events.

Similar sets of correlation functions were generated as a function of ${\pT}$ 
and  $\langle \mathrm{N_{ch}} \rangle$ to allow a study of  ${v^{\text{even}}_{1}}$, 
${v_2}$ and ${v_3}$  (for each collision system) for different 
dimensionless sizes and eccentricities. Monte Carlo quark Glauber (MC-qGlauber) calculations \cite{Liu:2018hjh} 
were used to compute ${\varepsilon_n}$ as a function of collision centrality 
or $\langle \mathrm{N_{ch}} \rangle$ for all collision systems, from the two-dimensional profile of the density of 
quark participants in the transverse plane (c.f Eq.~\ref{epsdef1}).
The model takes account of the finite size of the nucleon, the wounding profile of the nucleon,
the distribution of quarks inside the nucleon, and quark cross sections which reproduce 
the NN inelastic cross section at $\sqrt{s_{NN}}$ = 200 GeV.

The ${v_{nn}}$ coefficients were obtained from the correlation function as:
\begin{eqnarray}\label{vn}
 {v_{nn}} &=& \frac{\sum_{\Delta\phi} C_{r}(\Delta\phi, \Delta\eta){\cos(n \Delta\phi)}}{\sum_{\Delta\phi}~C_{r}(\Delta\phi, \Delta\eta)},
\end{eqnarray}
and then used to extract ${v_{n}}$ for ${n >1}$,
\begin{eqnarray}
\label{eq:3}
{v_{nn}(\pT^{a},\pT^{b})  = v_n(\pT^{a})v_n(\pT^{b})},
\end{eqnarray}
and the ${v^{even}_{1}}$ component of $v_1$,
\begin{equation}
\label{corrv1}
{v_{11}(\pT^{a},\pT^{b})  = v^{even}_{1}(\pT^{a})v^{even}_{1}(\pT^{b})} - K {\pT^{a}\pT^{b}},
\end{equation} 
where $K \propto 1/(\langle \mathrm{N_{ch} }\rangle \langle \pT^{2}\rangle)$ takes account of the long-range non-flow correlations induced by global momentum conservation~\cite{Retinskaya:2012ky,ATLAS:2012at,Star:2018zpt}. A simultaneous fit of ${v_{11}(\pT^{b})}$ for several selections of ${\pT^{a}}$ (c.f. Eq.~\ref{corrv1}) was used to facilitate the extraction of ${v^{\text{even}}_{1}}$ \cite{Star:2018zpt}.

The systematic uncertainties associated with the ${v_n}$ extractions were estimated through studies of the influence of  
the choice of the cuts for z-vertex position, track selection, efficiency correction, $\Delta\eta$ and the 
fitting procedure.  The uncertainty associated with $\Delta\eta$ dominate for the $d$+Au and $p$+Au systems.
The respective uncertainties, ranging from 2\% to 10\%, were added in quadrature to obtain an overall systematic 
uncertainty for the respective measurements.
%
\begin{figure*}[t]
\centering{
\includegraphics[width=0.65\linewidth,angle=0.0]{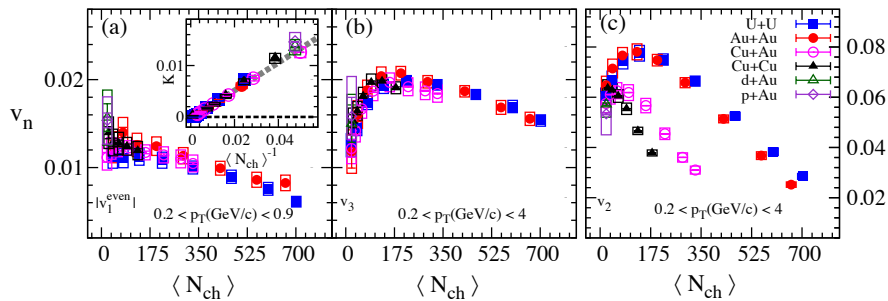}
\vskip -0.35cm
\caption{Comparison of the $\langle \mathrm{N_{ch}} \rangle$ dependence of 
${v^{\text{\text{even}}}_{1}}$ (a), $v_3$ (b) and ${v_2}$ (c) for all 
collision systems for the ${\pT}$ selections indicated. The $\langle \mathrm{N_{ch}} \rangle$
values for $p$+Au and $d$+Au correspond to $\sim$~0-20\% central collisions. The inset in (a) compares 
the extracted values of $K$ vs. $\langle \mathrm{N_{ch}} \rangle^{-1}$ for each system; the dashed 
line is drawn to guide the eye.
 \label{vn:mult}
 }
}
\vskip -0.45cm
\end{figure*}
%


The extracted values of ${v^{\text{\text{even}}}_{1}(\pT)}$, ${v_2(\pT)}$ and 
${v_3(\pT)}$ for the collision systems are compared  in Fig.~\ref{vn:pt} for 
different values of $\langle \mathrm{N_{ch}} \rangle$. Figures~\ref{vn:pt}(a) - \ref{vn:pt}(c) indicate 
similar ${v^{\text{even}}_{1}(\pT)}$ magnitudes for the systems 
specified at each $\langle \mathrm{N_{ch}} \rangle$, as well as the characteristic pattern of 
a change from negative ${v^{\text{even}}_{1}(\pT)}$ at low-${\pT}$ to positive 
${v^{\text{even}}_{1}(\pT)}$ for ${\pT \agt 1}$~GeV/$c$. This pattern confirms the predicted 
trends for dipolar flow \cite{Teaney:2010vd,Luzum:2010fb,Retinskaya:2012ky,Star:2018zpt} 
and further indicates that for the selected values of $\langle \mathrm{N_{ch}} \rangle$,  
${v^{\text{even}}_{1}(\pT)}$ is essentially independent of collision system. 
Figures~\ref{vn:pt}(d) - \ref{vn:pt}(f), show similar system-independent 
patterns for ${v_3(\pT)}$, but with magnitudes and trends that differ from those 
for ${v^{\text{even}}_{1}(\pT)}$. The system independence of ${v^{\text{even}}_{1}(\pT)}$ 
and ${v_3(\pT)}$ for the indicated $\langle \mathrm{N_{ch}} \rangle$ values suggests
that the fluctuations-driven initial-state eccentricities $\varepsilon_1$ and $\varepsilon_3$, 
and the subsequent final-state interactions are similar for the indicated collision systems.

The ${v_2(\pT)}$ values shown in  Figs.~\ref{vn:pt}(g) - \ref{vn:pt}(i) contrasts with those for ${v^{\text{even}}_{1}(\pT)}$ and  ${v_3(\pT)}$. That is, the trends for a given $\langle \mathrm{N_{ch}} \rangle$ 
are independent of the collision system, but the magnitudes are not system-independent, albeit with differences that grow with  $\langle \mathrm{N_{ch}} \rangle$.
 The system dependent differences, apparent for $\langle \mathrm{N_{ch}} \rangle$ = 140 and 70 (Figs.~\ref{vn:pt}(g) and \ref{vn:pt}(h)), 
can be attributed to the system-dependent $\varepsilon_2$ values for each $\langle \mathrm{N_{ch}} \rangle$. 
For $\langle \mathrm{N_{ch}} \rangle \sim 21$ (Fig.~\ref{vn:pt}(i)), the MC-qGlauber eccentricities for the 
different systems, do not vary strongly.

Figures~\ref{vn:pt}(j) and \ref{vn:pt}(k) confirm the influence of the system-dependent  
$\varepsilon_2$ values for $\langle \mathrm{N_{ch}} \rangle$ = 140 and 70 . That is, they show data 
collapse onto a single curve for ${v_{2}/\varepsilon_{2}}$ vs. $\pT$ 
for U+U, Au+Au, Cu+Au and Cu+Cu systems.  Fig.~\ref{vn:pt}(l) also indicates an approximate collapse of the scaled 
results for $p$+Au and $d$+Au onto the curve for the eccentricity-scaled A+A data.
This pattern is suggestive of a dominant collective flow contribution to the measured anisotropy in high 
multiplicity $p$+A($d$+A) collisions \cite{Liu:2018xae}. However, a quantitative estimate of a possible long-range non-flow 
contribution is required to fully establish the degree of this apparent scaling.

The $\langle \mathrm{N_{ch}} \rangle$ dependence of  ${v^{\text{even}}_{1}}$, ${v_2}$, and ${v_3}$ 
are compared  for all six collision systems in  Figs.~\ref{vn:mult}(a) - \ref{vn:mult}(c); they are in good 
agreement with the $v_2$ data reported for U+U and Au+Au collisions in Ref. \cite{Adamczyk:2015obl}. 
The inset in Fig.~\ref{vn:mult}(a) compares the associated values of $K$ vs. $\langle \mathrm{N_{ch}} \rangle^{-1}$ 
(c.f. Eq.~\ref{corrv1}) for each system. 

For $\langle \mathrm{N_{ch}} \rangle \agt 170$, the ${v_n}$ 
values all show a decrease with increasing values of $\langle \mathrm{N_{ch}} \rangle$, consistent with the 
expected decrease of ${\varepsilon_n}$ as collisions become more central. The apparent decrease in 
the values of ${v_{2}}$ for $\langle \mathrm{N_{ch}} \rangle \alt 170$  corroborate the dominant role 
of size-driven viscous attenuation of the flow harmonics for these multiplicities. Note that $\varepsilon_2$ 
increases for $\langle \mathrm{N_{ch}} \rangle \lt 170$.  Figures~\ref{vn:mult}(a) and \ref{vn:mult}(b) 
indicate system-independent magnitudes and trends for ${v^{\text{even}}_{1}}$ and ${v_3}$ analogous 
to the ${\pT}$-dependent results shown in Fig.~\ref{vn:pt}. 

\begin{figure}
\centering{
\includegraphics[width=0.6\linewidth,angle=-90]{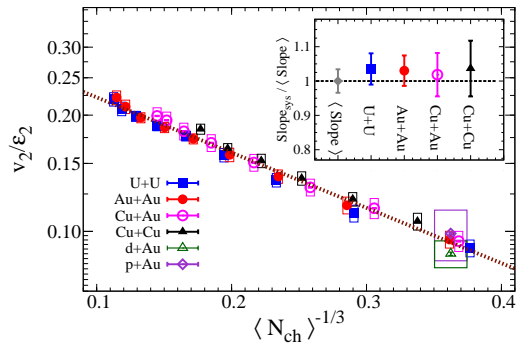}
\vskip -0.66cm
\caption{ ${v_2/\varepsilon_2}$ vs. $\langle \mathrm{N_{ch}} \rangle^{-1/3}$ 
for U+U, Au+Au, Cu+Au, Cu+Cu, $d$+Au and $p$+Au collisions as indicated. The open boxes indicate systematic uncertainties. The ${v_2}$ 
data is the same as in Fig. \ref{vn:mult}(c). The dotted line represents an exponential fit to the data with Eq.~\ref{eq:4}. 
The inset shows the respective ratios of the slopes extracted for each system relative to the 
slope extracted from a fit to the combined data sets ($\langle \mathrm{Slope} \rangle$  = $8.2 \times 10^{-1}\pm 0.02$).
\label{v2:mult}
 }
}
\end{figure}

The ${v_2}$ comparisons shown in Fig.~\ref{vn:mult}(c), accentuate the system-dependent patterns 
observed in Figs.~\ref{vn:pt}(g), \ref{vn:pt}(h) and \ref{vn:pt}(i). Here, the sizable uncertainties for 
the $p$+Au and $d$+Au data points for $\langle \mathrm{N_{ch}} \rangle \sim 21$ reflect the systematic uncertainty estimates 
for residual non-flow contributions which are smaller for these $\pT$-integrated measurements. The striking system-dependent 
patterns shown in Fig.~\ref{vn:mult}(c) can be attributed to the strong dependence of $\varepsilon_2$ on 
system size for a fixed value of $\langle \mathrm{N_{ch}} \rangle$. This shape dependence, which weakens 
for low $\langle \mathrm{N_{ch}} \rangle$, is confirmed via the plot 
of ${v_{2}/\varepsilon_{2}}$ vs. $\langle \mathrm{N_{ch}} \rangle^{-1/3}$ shown 
in Fig.~\ref{v2:mult}.  A similar plot, reflecting the $n^2$ dependence of viscous attenuation \cite{Liu:2018hjh,Liu:2018xae}, was obtained
for ${v_{3}/\varepsilon_{3}}$ vs. $\langle \mathrm{N_{ch}} \rangle^{-1/3}$.
The inset in Fig.~\ref{v2:mult} indicates a marked similarity between the slopes of the 
eccentricity-scaled ${v_2}$ for U+U, Au+Au, Cu+Au and Cu+Cu collisions. 
The eccentricity-scaled results for $d$+Au and $p$+Au also follow the data trend for these heavier collision species with 
larger systematic uncertainty.  Hydrodynamic simulations for Au+Au collisions \cite{Alba:2017hhe}
exhibit similar scaling trends within the same range of $\langle \mathrm{N_{ch}} \rangle$.
%
%

In summary, we have used the two-particle correlation method to carry out a comprehensive set of measurements of ${v^{\text{even}}_{1}}$, ${v_2}$, and ${v_3}$ as a function 
of ${\pT}$ and $\langle \mathrm{N_{ch}} \rangle$ in U+U ($\roots= 193$~GeV) and Au+Au, Cu+Au, Cu+Cu, $d$+Au, and $p$+Au  
collisions at $\roots = 200$~GeV.  The detailed comparisons of the measurements highlight the sensitivity of ${v_n}$ to 
the magnitude of the initial-state eccentricity, system size and the final-state interactions in the expanding matter. 
The wealth of the A+A measurements 
lead to data collapse of  ${\ln(v_{n}/\varepsilon_{n})}$ vs. $\langle \mathrm{N_{ch}} \rangle^{-1/3}$ onto a single curve.
Similarly scaled results for $d$+Au and $p$+Au (for $\langle \mathrm{N_{ch}} \rangle \sim 21$) are also observed with larger uncertainty.
The combined measurements and their scaling properties provide a new set of constraints which could prove invaluable 
for the interpretation of collectivity in small systems and for detailed theoretical extraction of the temperature-dependent $\frac{\eta}{s}$. 

\section*{Acknowledgments}
\begin{acknowledgements}
%
We thank the RHIC Operations Group and RCF at BNL, the NERSC Center at LBNL, and the Open Science Grid consortium for providing resources and support. 
This work was supported in part by the Office of Nuclear Physics within the U.S. DOE Office of Science, the U.S. National Science Foundation, the Ministry 
of Education and Science of the Russian Federation, National Natural Science Foundation of China, Chinese Academy of Science, the Ministry of Science 
and Technology of China and the Chinese Ministry of Education, the National Research Foundation of Korea, GA and MSMT of the Czech Republic, 
Department of Atomic Energy and Department of Science and Technology of the Government of India; the National Science Centre of Poland, 
National Research Foundation, the Ministry of Science, Education and Sports of the Republic of Croatia, RosAtom of Russia and German 
Bundesministerium fur Bildung, Wissenschaft, Forschung and Technologie (BMBF) and the Helmholtz Association.
%
\end{acknowledgements}
%
%
\bibliography{ref_vn_diff_sys} 
\end{document}